\documentclass[twocolumn,aps,epsfig,showpacs]{revtex4}
\usepackage{graphicx}  
\newcommand{\3}{$^3$He}

\newcommand{\4}{$^4$He}
\begin{document}
\title{Dissipation of Quantum Turbulence in the Zero Temperature Limit}
\author{P. M. Walmsley$^a$, A. I. Golov$^a$, H. E. Hall$^a$, A. A. Levchenko$^b$, and W. F. Vinen$^c$ }
\address{$^a$School of Physics and Astronomy, The University of Manchester, Manchester M13 9PL, UK \\$^b$Institute of Solid State Physics, Russian Academy of Sciences,  Chernogolovka 142432, Russia \\$^c$School of Physics and Astronomy, University of Birmingham, Birmingham B15 2TT, UK}
\date{\today}
\begin{abstract}
Turbulence, produced by an impulsive spin-down from angular velocity $\Omega$ to rest of a cube-shaped container, is investigated in superfluid $^4$He at temperatures 0.08~K -- 1.6~K. The density of quantized vortex lines $L$ is measured by scattering negative ions. Homogeneous turbulence develops after time $t \approx 20/\Omega$ and decays as $L \propto t^{-3/2}$. The corresponding energy flux $\epsilon = \nu' (\kappa L)^2 \propto t^{-3}$ is characteristic of quasi-classical turbulence at high Re with a saturated energy-containing length. The effective kinematic viscosity in $T=0$ limit is $\nu' = 0.003 \kappa$, where $\kappa=10^{-3}$~cm$^2$s$^{-1}$ is the circulation quantum. 
\end{abstract}
\maketitle
We report the first measurements of the decay of homogeneous turbulence in superfluid $^4$He in the zero temperature limit, where thermal excitations are effectively absent. The turbulence,  produced by bringing a cubical container rapidly to rest from a state of rotational equilibrium, has the classical Kolmogorov form on large length scales,  even though on short scales rotational motion is severely restricted by the quantization of circulation. At $T<0.8$~K, we observe a drop in the effective kinematic viscosity,  by a factor of $\sim 30$,  to an ultimate $T=0$ value of $0.003 \kappa$.  We associate this drop with a regime in which the energy flux in the Kolmogorov cascade, when it reaches the quantum scale, must be transferred to a Kelvin-wave cascade in which energy is carried to much smaller, dissipative,  length scales  by Kelvin waves on the individual quantized vortex lines.

Superfluid  \4 (strictly the superfluid component) is inviscid, and rotational flow can be achieved only with line defects, {\it quantized vortices} \cite{Donnelly1991}, each carrying one quantum of irrotational circulation $\kappa \equiv h/m$ outside a core of radius $a_{\rm 0} \sim 1$~\AA. Turbulence in the superfluid (``quantum turbulence'') can exist as a dynamic tangle of these vortices \cite{VinenJLTP2006}, characterized by their total length per unit volume, $L$. On length scales much larger than the average distance between vortices, $\ell \equiv L^{-1/2}$, rotational flow is possible as the result of partial alignment of the vortices \cite{HallVinen1957,Vinen2000}. Hence, at sufficiently large scales the superfluid should be able to support quasi-classical turbulence, similar to that in classical liquids at large Reynolds numbers,  in which the energy cascades  towards smaller eddies until it is dissipated \cite{Frisch}. Numerical simulations \cite{Araki2002} confirm this. Dissipation is through the interaction of quantized vortices with excitations in the liquid. In the zero-temperature limit emission of phonons by Kelvin waves at wavelengths $\sim 100$~\AA\ is expected \cite{Vinen2000, KS2004}, while above $T \approx 0.5$~K, scattering of the thermal excitations (``mutual friction'') should dominate \cite{HallVinen1957}.

We see that in the $T=0$ limit there is a need to transfer energy at scales of order $\ell$ from quasi-classical turbulent flow to ultra-quantum Kelvin-wave turbulence. L'vov {\it et al.} \cite{LNR2007} have suggested that this transfer involves a bottleneck,  so that continuity of the energy flux requires that the vortex line density at mesoscales $\sim \ell$ should be enhanced. Kozik and Svistunov \cite{KS2007} argue that reconnections operating in a range of wave numbers, covering an intermediate regime between the Kolmogorov spectrum and the Kelvin-wave spectrum, allow the bottleneck to be effectively bypassed.

The rate of dissipation, mediated by vortex lines, per unit mass can be written in the form 
\begin{equation}
\epsilon = \nu' (\kappa L)^2,
\label{epsilon}
\end{equation}
 where $\kappa^2 L^2$ is an effective total mean square vorticity \cite{Vinen2000} and the parameter $\nu'$ is an ``effective kinematic viscosity''.

In previous studies of quasi-classical turbulence in superfluid \4 at $T>1$~K, turbulence that was nearly isotropic and homogeneous was generated by either rotating  blades \cite{MT1998} or a towed grid \cite{Stalp1999}.  To extract $\nu'$ \cite{Stalp1999}, one can monitor the late-time decay of the vortex density,  which has the form $L(t)\propto \nu'^{-1/2}t^{-3/2}$. The following assumptions lead to this type of decay. The turbulent energy should be concentrated in the largest eddies characterized by a size $d$ and velocity $u$,  so that the total turbulent energy per unit mass is $E\approx u^2/2$; $d$ must be constant in time,  as is the case if it is limited by the container size;  and the lifetime of the largest eddies with respect to breaking up into smaller eddies without dissipation is of order the turn-over time $\tau \sim d/u$. Hence, the energy flux $\epsilon = -\dot{E} \sim u^3/d \sim - E^{3/2}/d$. This yields $E \sim d^2(t+t^*)^{-2}$ and $\epsilon \sim d^2(t+t^* )^{-3}$, where $t=0$ coincides with the time of activation. Here, $t^*$ is an arbitrary constant absorbing the duration of the transient processes and the initial value of the turbulent energy. It is usually of order $\tau$, and hence in what follows we will omit it as we are interested in the late-time decay for $t \gg \tau$. Finally, using Eq.~(1) we obtain $\kappa L \sim d \nu'^{-1/2} t^{-3/2}$. We shall use this relation with the numerical prefactor,    
\begin{equation}
L = (3C)^{3/2} (2\pi\kappa)^{-1} d \nu'^{-1/2} t^{-3/2},
\end{equation}
derived  by Stalp {\it et al.} \cite{Stalp1999, Stalp2002} for homogeneous isotropic turbulence in a channel of square ($d\times d$) cross-section,  using the Kolmogorov energy spectrum with $C = 1.5$, 
\begin{equation}
E_k = C\epsilon^{2/3}k^{-5/3}
\label{Ek}
\end{equation}
and with an assumed sharp cut-off for $k < 2\pi/d$.

Bradley {\it et al.} \cite{Bradley} have measured $L(t)$ produced by a vibrating grid in \3-B down to $T=0$ limit, and observed a free decay $L(t) \propto t^{-3/2}$. By assuming a value of $d$ in Eq.~(1) equal to the measured spatial spread of the turbulence, they inferred $\nu' \sim 0.2 \kappa$. While their observations are interesting and suggestive, the turbulence is not homogeneous,  and the size of the energy-containing eddies may differ from the spatial extent of the turbulence, so that the value of $\nu'$ is ambiguous. Other experiments on the dynamics of inhomogeneous turbulence in \4 (vibrating grid) \cite{McClintock} and \3-B (propagating turbulent front)  \cite{Eltsov} hinted at new dissipation mechanisms in $T=0$ limit too. 

The need for experiments with homogeneous turbulence in superfluid \4 generated by large-scale flow in $T=0$ limit has led us to an experiment in which energy-containing eddies comparable in size to that of the container are generated by an impulsive spin-down to rest of a rotating cell of a square cross-section. It is known that with classical fluids spin-down to rest is always unstable resulting in 3-dimensional turbulence after a few radians of the initial rotation \cite{vanHeijst1989}. The values of $L$ in our experiments, monitored by firing a beam of charged probes of known trapping diameter through the tangle, were in the range $10^1$--$10^5$~cm$^{-2}$, corresponding to characteristic inter-vortex distances $\ell$ of 3 -- 0.03~mm.

\begin{figure}[h]
\centerline{\includegraphics[width=8.5cm]{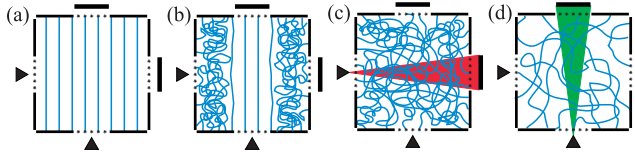}}
\caption{(color online) Cartoon of the vortex configurations in the experimental cell (side view) at different stages. (a) Regular array of vortex lines during rotation at constant $\Omega$ before deceleration. (b) Immediately after stopping rotation ($0 < t < 10/\Omega$), turbulence appears at the outer edges but not on the axis of rotation. (c) After about $t\sim 30/\Omega$, 3D homogeneous turbulence is everywhere. (d) At $t \sim 10^3/\Omega$  most of the 3D turbulence has decayed. Shaded areas indicate the paths of probe ions when sampling the vortex density in the transverse ($L_{\rm t}$, c, red) and axial ($L_{\rm a}$, d, green) directions.} \label{fig1}
\end{figure}

The inner experimental volume of the container had the shape of a cube of 4.5~cm side;  it was filled with pure \4 at a pressure of 0.1~bar, and mounted on a rotating cryostat  (Fig.~1; for details see \cite{IonCellPobell}). Negative ions could be injected by either of the two tungsten field-emission tips \cite{IonTip} through gridded holes in the centers of the bottom and of one of the side plates. Then the ions could be pulled through the volume by an applied electric field $E$ (converging to counteract the space-charge repulsion) and collected by the electrodes in the center of the opposite plates protected by Frisch grids. To make sure the free decay of the tangle and measurements of $L$ are not affected by the space charge, only one short pulse (duration $\leq 0.5$~s) of injected ions was fired, at time $t$ after stopping rotation. Thus each data point in Figs.~2--4 represents a different realization of the turbulence. Before decelerating to rest, the cryostat was kept in steady rotation at the required $\Omega$ for at least 300~s. The deceleration was linear in time,  taking 2.5~s for $\Omega = 1.5$~rad/s and 0.1~s for $\Omega = 0.05$~rad/s. The origin $t=0$ was chosen at the start of deceleration. 

\begin{figure}[h]
\centerline{\includegraphics[width=8cm]{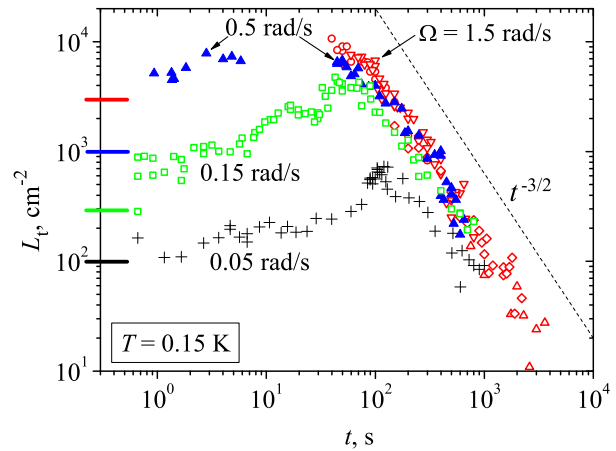}}
\caption{(color online) $L_{\rm t}(t)$ at $T=0.15$~K for four values of $\Omega$. Average electric fields used for $\Omega = 1.5$~rad/s: 5~V/cm ($\diamond$), 10~V/cm ($\bigtriangleup$), 20~V/cm ($\circ$), 25~V/cm ($\bigtriangledown$). The dashed line shows the dependence $t^{-3/2}$. Horizontal bars indicate the equilibrium values of $L$ at $\Omega = 1.5$, 0.5, 0.15. 0.05~rad/s (from top to bottom).} \label{fig2}
\end{figure}

The pulse of collector current arrives after a well-defined time of flight across the cell,  and the relative reduction in its amplitude,  $I(t)/I(\infty)$, is converted into the average vortex density through $L(t)/L_0 = (\sigma d)^{-1}\ln(I(\infty)/I(t))$. There are two types of charge carrier \cite{Donnelly1991}. Free ions, dominating at $T>0.8$~K, are electron-containing bubbles of radius $\approx 19$~\AA; at $T=1.6$~K their trapping diameter $\sigma \propto E^{-1}$ was 0.1~$\mu$m at $E=20$~V/cm, and the time of flight  was 1~s. The others, dominating at $T < 0.7$~K, are small quantized vortex rings of diameter $D \sim 1$~$\mu$m with one electron bubble trapped in its core; at $T=0.15$~K their trapping diameter $\sigma \sim D$ was between 0.4 and 1.7~$\mu$m for $E$ between 5 and 25~V/cm.  In the narrow range of temperatures, 0.7 -- 0.8~K, both types can exist, distinguishable at sufficiently high fields because the free ions move much faster than the rings; here measurements of $L$ with both types of carriers agreed well. Their trapping diameters $\sigma$ and times of flight were calibrated {\it in situ}, for all temperatures and field strengths used, on arrays of parallel rectilinear vortex lines with a range of equilibrium densities $L=2\Omega/\kappa$ produced by steady rotation at different angular velocities $\Omega$. They agreed with the published data \cite{SchwarzDonnelly1966, OstermeierGlaberson1974}. We could not measure values of $L$ that are too high (no detectable current reaches the collector) or too low (no detectable change in current). For charged vortex rings and free ions, the accessible ranges of $L$ were about $10^1$--$10^4$ and $5\cdot10^2$--$10^6$~cm$^{-2}$, respectively. At low temperatures, with Kelvin waves excited across a range of wave numbers, the vortex length is a fractal property, and hence can be underestimated by the use of probe objects of size $\sigma \sim 1$~$\mu$m. However, as the values of $L$ at $T=0.15$~K, measured for a range of $\sigma = 0.4$--1.7~$\mu$m, were in good agreement, any undetected contribution to $L$ must be small.   In Fig.~2, the measured densities of vortex lines along the horizontal axis (transverse, $L_{\rm t}$) are shown for four different initial angular velocities $\Omega$. During the transient, which lasts $\sim 10/\Omega$, $L_{\rm t} (t)$ goes through a maximum,  after which it decays,  eventially reaching the universal late-time form of $L \propto t^{-3/2}$. 

\begin{figure}[h]
\begin{center}
\includegraphics[width=8cm]{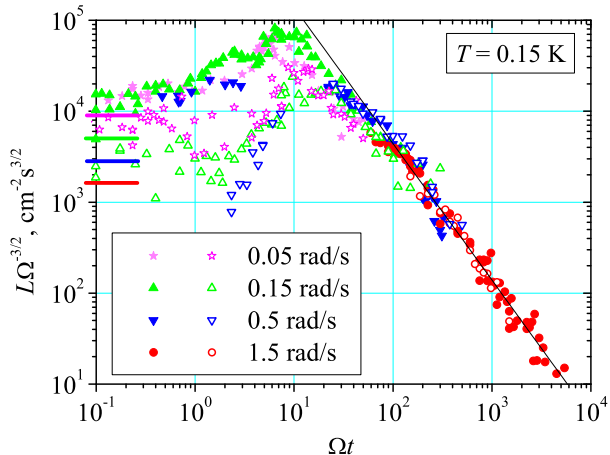}
\caption{(color online) $\Omega^{-3/2}L_{\rm t}(t)$ (filled symbols) and $\Omega^{-3/2}L_{\rm a}(t)$ (open symbols) vs. $\Omega t$ for four values of $\Omega$ at $T=0.15$~K. The straight line $\propto t^{-3/2}$ guides the eye. Horizontal bars indicate the equilibrium values of $L$ at $\Omega = 0.05$, 0.15, 0.5. 1.5~rad/s (from top to bottom).} 
\label{fig3}
\end{center}
\end{figure}

In Fig.~3 the measured densities of vortex lines along the horizontal, $L_{\rm t}$, and vertical (axial, $L_{\rm a}$) axes are shown, by solid and open symbols respectively. To stress the scaling of the characteristic times with the initial turn-over time $\Omega^{-1}$ and the universal late-time decay $\propto t^{-3/2}$, the data for different $\Omega$ are rescaled accordingly. At all $\Omega$ the transients are universal. Immediately after deceleration, $L_{\rm a}$ is stable near the equilibrium density at rotation, $2\Omega/\kappa$, while $L_{\rm t}$ jumps above this value, indicating the appearance of turbulence at the perimeter. Only at $t \approx 3/\Omega$ does the former start to grow, signalling the destruction of the rotating core consisting of vertical rectilinear vortices. After passing through a maximum at $t = 8/\Omega $ and $t = 15/\Omega $ respectively, $L_{\rm t}$ and $L_{\rm a}$ merge at $t \sim 30/\Omega$ and then become indistinguishible. This implies that from then on the turbulence in the cell becomes isotropic and probably homogeneous. The scaling of the transient times with the turn-over time $\Omega^{-1}$ tells us that transient flows are similar at different initial velocities $\Omega$, as expected for flow instabilities in classical liquids. 

\begin{figure}[h]
\begin{center}
\includegraphics[width=8cm]{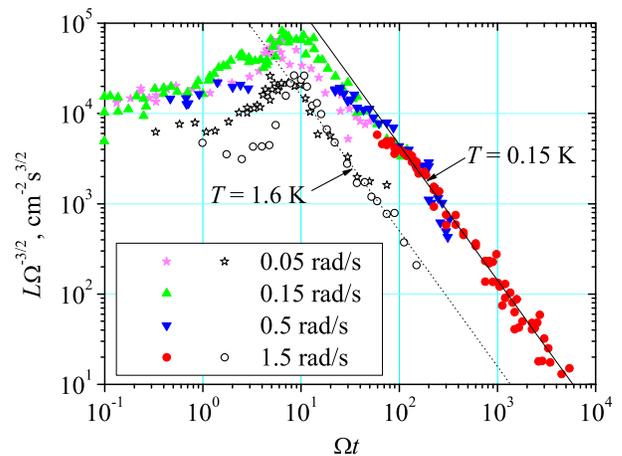}
\caption{(color online) $\Omega^{-3/2}L_t(t)$ vs. $\Omega t$ for $T=0.15$~K (filled symbols) and $T=1.6$~K (open symbols). Dashed and solid lines $\propto t^{-3/2}$ guide the eye at $T=1.6$~K and 0.15~K, respectively.} \label{fig4}
\end{center}
\end{figure}
At $0.08<T<0.5$~K the observed $L(t)$ and hence inferred $\nu'(T)$ were independent of temperature. In Fig.~4, we compare the transients $L_{\rm t}(t)$ at low (0.15~K) and high (1.6~K) temperatures. Their durations (positions of the maximum in $L(t)$) are virtually identical. On the other hand, at $T=0.15$~K, the late-time decay $L\propto t^{-3/2}$ takes longer  to develop (time $\sim 100/\Omega$) and then the prefactor is about 10 times larger than for $T=1.6$~K. This implies that at low temperatures the steady-state inertial cascade with a constant energy-containing eddy size and constant energy flux down the range of length scales  requires a much greater total vortex line density and perhaps some extra time to build up this density. 
\begin{figure}[h]
\begin{center}
\includegraphics[width=7.5cm]{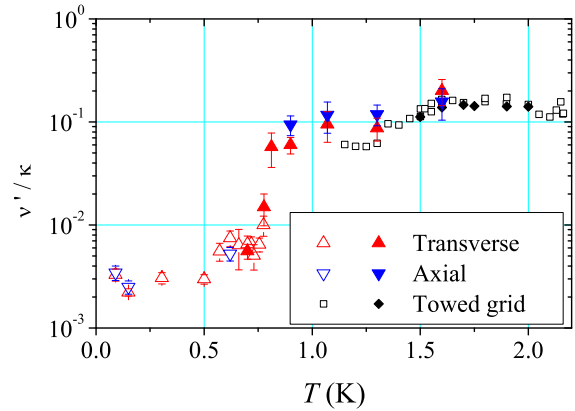}
\caption{(color online) The effective kinematic viscosity $\nu'$ after a spin-down from $\Omega=1.5$~rad/s measured in the transverse ($\bigtriangleup$) and axial ($\bigtriangledown$) directions. Closed (open) triangles correspond to measurements with free ions (charged vortex rings). Error bars specify the uncertainty of fitting. Squares and diamonds: second sound measurements of grid turbulence \cite{Stalp2002, Niemela2005}.}
\label{fig5}
\end{center}
\end{figure}

Our values of $\nu'$, along with those obtained with towed grids of two different designs in a channel of $1\times1$~cm$^2$ cross-section at high temperatures (corrected by the factor of 0.72 as suggested in \cite{SkrbekPRE2007}), are shown in Fig.~5. In discussing the behaviour of $\nu'(T)$, we note first that in the low temperature limit a quasi-classical inertial-range energy spectrum cannot persist to wave numbers greater in order of magnitude than $\pi/\ell$.  At larger wave numbers quantum effects must dominate. Therefore the maximum vorticity associated with the quasi-classical spectrum cannot be greater than that given by
\begin{equation} \langle\omega^2\rangle_{\mathrm{max}}^{\mathrm{(class)}} \approx \int_{0}^{\pi/\ell} k^2 E_k dk = \frac{3C}{4} \pi^{4/3}\epsilon^{2/3} L^{2/3}.
\label{max1}
\end{equation} 

The energy flux $\epsilon$ is given by Eq.~(\ref{epsilon}), therefore
\begin{equation} \frac{\langle\omega^2\rangle_{\mathrm{max}}^{\mathrm{(class)}}}{\kappa^2 L^2}=\frac{3C}{4} \pi^{4/3} \Big( \frac{\nu'}{\kappa}\Big)^{2/3} = 5.2\Big(\frac{\nu'}{\kappa}\Big)^{2/3}.
\label{max3}
\end{equation}

Now $\kappa^2 L^2$ can be identified as an effective total mean square vorticity in the system \cite{Vinen2000}.  Thus, if $\nu'/\kappa <0.09$ the quasi-classical spectrum cannot account for all the vorticity in the system.  In the low-temperature limit some,  but not,  according to our estimates, all,  of the extra vorticity could be associated with the Kelvin-wave cascade;  some must exist either at the high-wavenumber end of the quasi-classical cascade,  associated with a bottleneck,  or in an intermediate regime of the type described by Kozik and Svistunov \cite{KS2007}. 
According to the detailed calculations of L'vov \textit{et al.} \cite{LNR2007} the ratio $\nu'/\kappa$ is only about $(\ln(\ell/ a_0))^{-5}=10^{-6}$,  and therefore our measurements suggest that any bottleneck is less severe than these calculations predict.  We remark that, although there is vorticity in excess of that given by the Kolmogorov cascade,  the associated energy must be small enough not to invalidate our assumption that the largest eddies account for most of the energy; otherwise the $t^{-3/2}$ decay would not be observed.  

The striking temperature dependence of $\nu'/\kappa$ shown in Fig.~5 therefore suggests to us the following scenario.  As the temperature falls from 1.3~K to 0.8~K the Kolmogorov cascade expands to reach wavenumbers of order $\pi/\ell$.  Below 0.8~K (when the estimated mutual friction parameter $\alpha \sim \rho_{\rm n}/\rho$ becomes $\alpha \leq 10^{-3}$) the flow of energy for a given line density drops sharply as vorticity accumulates at wavenumbers around $\pi/\ell$ and Kelvin waves begin to be excited.  As the temperature falls further the damping of Kelvin waves by mutual friction decreases until below 0.5~K the Kelvin-wave cascade extends to wavenumbers far beyond $\pi/\ell$.  The low value of $\nu'$ in the zero temperature limit is a measure of the difficulty in transferring energy through wavenumbers around $\pi/\ell$ from the 3-dimensional quasi-classical Kolmogorov cascade to the 1-dimensional Kelvin-wave cascade. Better understanding of this process remains a challenge for theorists.

We thank Steve May and Sio Lon Chan for their contribution to the construction and improvement of the apparatus. Support was provided by EPSRC under GR/R94855 and EP/E001009.

\end{document}